\documentclass[useAMS,usenatbib]{mn2e} 
\usepackage{times}

\def\spose#1{\hbox to 0pt{#1\hss}} 
\newcommand\lsim{\mathrel{\spose{\lower 3pt\hbox{$\mathchar"218$}} 
     \raise 2.0pt\hbox{$\mathchar"13C$}}} 
\newcommand\gsim{\mathrel{\spose{\lower 3pt\hbox{$\mathchar"218$}} 
     \raise 2.0pt\hbox{$\mathchar"13E$}}} 
 
\title[The $L_{\rm iso}$--$E_{\rm peak}$ correlation]
{The peak luminosity -- peak energy correlation in GRBs} 
 
\author[Ghirlanda G. et al.]  
{
G. Ghirlanda$^{1}$\thanks{E-mail: ghirlanda@merate.mi.astro.it}, 
G. Ghisellini$^{1}$,  
C. Firmani$^{1,2}$,  
A. Celotti$^{3}$, 
Z. Bosnjak$^{3}$  \\ 
$^{1}$Osservatorio Astronomico  di Brera, via E.Bianchi 46, 
   I--23807   Merate,   Italy\\    
$^{2}$Instituto   de Astronom\'{\i}a,  U.N.A.M.,  A.P.    
  70-264, 04510,  M\'exico,  D.F., M\'exico\\ 
$^3$SISSA/ISAS, via Beirut 4, I--34014 Trieste, Italy
}

\begin{document} 
 
\date{} 
 
\pagerange{\pageref{firstpage}--\pageref{lastpage}} \pubyear{2002} 
 
\maketitle 
 
\label{firstpage} 
 
\begin{abstract} 
We derive the peak luminosity--peak energy 
($L_{\rm iso}$--$E_{\rm peak}$) correlation using 22 long 
Gamma--Ray Bursts (GRBs) with firm redshift measurements.  
We find that its slope is similar to the
correlation between the time integrated isotropic emitted energy
$E_{\rm iso}$ and $E_{\rm peak}$ (Amati et al. 2002).  For the 15 GRBs
in our sample with estimated jet opening angle we compute the
collimation corrected peak luminosity $L_\gamma$, and find that it
correlates with $E_{\rm peak}$. This has, however, a scatter larger than
the correlation between $E_{\rm peak}$ and $E_{\gamma}$ (the time
integrated emitted energy, corrected for collimation; Ghirlanda et
al. 2004), which we ascribe to the fact that the opening angle is
estimated through the global energetics.  We have then selected a
large sample of 442 GRBs with pseudo--redshifts, derived through the
lag--luminosity relation, to test the existence of the 
$L_{\rm iso}$--$E_{\rm peak}$ correlation.  
With this sample we also explore the possibility of a correlation 
between time resolved quantities, namely $L_{\rm iso}^{\rm p}$ and the peak 
energy at  the peak of emission $E_{\rm peak}^{\rm p}$.  
\end{abstract} 

 
\section{Introduction} 
 
Several  correlations   have  been  identified   among  the  intrinsic
properties of  the (small) population of GRBs  with measured redshifts
$z$.   In  particular two  spectral  correlations  have been  recently
discussed in the literature:  i) the ``Amati correlation'' between the
energy $E_{\rm peak}$  where most of the emission  is radiated and the
total emitted  energy (isotropic  equivalent) $E_{\rm iso}$  (Amati et
al.   2002 -  A02 hereafter;  Lloyd  \& Ramirez--Ruiz  2002); ii)  the
``Ghirlanda   correlation''    between   $E_{\rm   peak}$    and   the
collimation--corrected  energy $E_{\gamma}$ (Ghirlanda,  Ghisellini \&
Lazzati 2004 - GGL04 hereafter).  It is important to notice that these
correlations refer to the {\it time integrated} spectral properties of
GRBs.   This is  true for  both $E_{\rm  peak}$ and  for  the spectral
indices required to calculate  the rest frame bolometric $E_{\rm iso}$
and $E_{\gamma}$.

However, time  resolved spectral analysis  of large samples  of bursts
(e.g.  Ford  et al.   1995; Preece et  al.  2000;  Ghirlanda, Celotti,
Ghisellini 2002)  have proved  that the GRB  spectrum evolves  in time
during the prompt emission phase.  The spectral evolution is different
among different GRBs  (e.g. Ford et al.  1995)  and not clearly linked
to other GRB  global parameters (e.g. duration, number  of peaks, peak
flux).  This spectral evolution may be revealing of the time variation
of the  parameters of the  radiative process(es) acting in  GRBs (e.g.
Liang \& Kargatis  1996) and/or of the relativistic  properties of the
emitting  outflow  (e.g.   Ryde  \&  Petrosian  2002).   In  order  to
understand the  origin of such  correlations it is thus  compelling to
determine   whether   they    are   representing   global   energetics
characteristics  or  they hold  for  and  are  dominated by  the  time
resolved  spectral  properties,  as  expected  if  determined  by  the
emission process(es).  One obvious possibility is to test them against
the  peak luminosity,  well  defined  for all  bursts  with known  $z$
(e.g. Liang, Dai \& Wu 2004).

This issue has been recently  considered by Yonetoku et al. 2004 (Y04,
hereafter).  With  a sample of  12 GRBs of  known $z$ they  found that
$E_{\rm peak}\propto L_{\rm iso}^{0.5}$.  This correlation appeared to
be tighter  (but with similar slope) than  the $E_{\rm peak}$--$E_{\rm
iso}$  correlation, as  originally found  by A02.   Note that  the Y04
analysis adopts  $E_{\rm peak}$ and  the spectral indices of  the time
integrated spectrum and not the spectral properties at the peak flux.

In this Letter we  first re--examine the $E_{\rm peak}$--$L_{\rm iso}$
correlation (i.e. the ``Yonetoku correlation") with an enlarged sample
of 22 GRBs with  spectroscopically measured $z$ and published spectral
properties.  For 15 out of these  22 GRBs we have an estimate of their
jet opening angle $\theta_{\rm j}$ (GGL04).  We can thus calculate the
collimation corrected  peak luminosity $L_\gamma$ and  verify if there
exists  the   equivalent  of  the  Ghirlanda   correlation  --  namely
$L_\gamma$  replacing $E_\gamma$ (Sec.  2).  Then  we consider  a much
larger   sample  of   442  GRBs   with  $z$   estimated   through  the
lag--luminosity  correlation  (Band, Norris  \&  Bonnel  2004 -  BNB04
hereafter) to  test if the  Yonetoku correlation still holds  for this
whole sample  (Sec. 3).  In Sec. 4,  by means of this  same sample, we
also  study the  relation between  $L_{\rm iso}^{\rm  p}$  and $E_{\rm
peak}^{\rm p}$,  (i.e. using  spectral parameters at  the peak  of the
flux), to check whether this  correlation is tighter than the Yonetoku
one and we discuss the differences  between the two.  We find that the
Ghirlanda  correlation has  a smaller  scatter than  the corresponding
$E_{\rm peak}$--$L_{\gamma}$  correlation.  We give  an interpretation
of this result in Sec. 5 and draw our conclusions in the final Sec. 6.

In  this  paper  we  adopt  a  standard  $\Lambda$CDM  cosmology  with
$\Omega_{\Lambda}=0.7$, $\Omega_{\rm M}=0.3$ and $h_0=0.7$.

\section{The $L_{\rm iso}$--$E_{\rm peak}$ correlation} 

The bolometric $\gamma$--ray luminosity can be defined once the prompt
emission spectrum and  the redshift $z$ of the  source are known.  GRB
spectra are typically described by the Band function $N(\alpha, \beta,
E_{\rm  peak})$ (Band  et al.  1993),  parameterized by  low and  high
energy   power--laws  (of   photon  indices   $\alpha$   and  $\beta$,
respectively) and by  peak energy $E_{\rm peak}$ in  the $\nu F_{\nu}$
representation.

The burst  emission varies on short timescales  (e.g. Ramirez--Ruiz \&
Fenimore  1999)  and  no  universal  temporal  profile  describes  the
``zoology'' of burst light curves (e.g. Norris et al. 1996).  However,
in most cases,  a dominating peak, with flux  $\Phi$ integrated in the
observed energy band,  can be identified in the  prompt emission light
curve.  The  rest frame,  bolometric (e.g.  1--10$^4$  keV), isotropic
peak  luminosity,  including  the  redshift--energy  band  correction,
follows straightforwardly.
%
%
%
\begin{figure} 
\begin{center} 
\vspace{8cm} 
\includegraphics{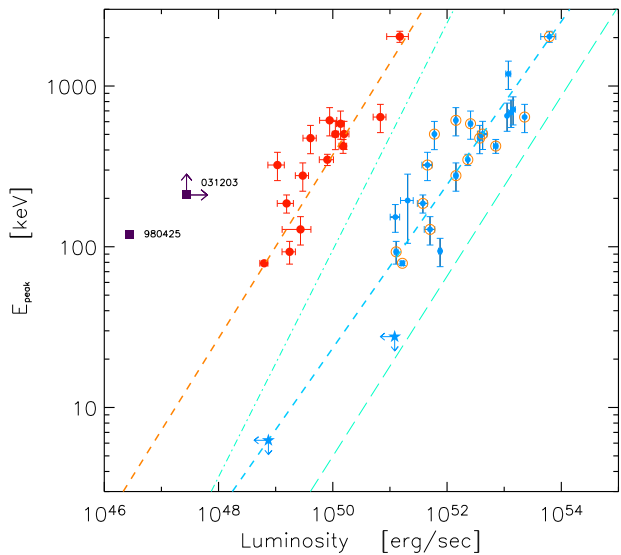}
\caption{Rest  frame  peak   energy  $E_{\rm  peak}=E_{\rm  peak}^{\rm
obs}(1+z)$ versus  the bolometric peak luminosity.  Samples: GRBs with
measured $z$ listed in Tab.~1 (blue symbols) -- upper/lower limits are
excluded except for the 2 X--ray Flashes (stars), shown for comparison
with Fig.~1  of GGL04; 15  GRBs with a  firm measure of the  jet break
time  (from  Tab.~2 of  GGL04)  and hence  of  the  jet opening  angle
$\theta_{\rm j}$  (open yellow circles);  same 15 GRBs  once corrected
for  the  $(1-\cos\theta_{\rm  j})$  collimation  factor  (red  filled
circles).  Fits: best  power--law fit  to the  $E_{\rm peak}$--$L_{\rm
iso}$   correlation    (dashed   blue   line);    best   fit   $E_{\rm
peak}$--$L_{\gamma}$   correlation  (dashed   red  line);   the  Amati
correlation from 23  GRBs in GGL04 and GGF05  (long--dashed blue line)
and the Ghirlanda correlation from GGL04 (dot--dashed blue line).  }
\end{center} 
\label{corr} 
\end{figure} 
In Tab.~1 we  report the peak luminosities of the  29 GRBs examined by
GGL04, computed assuming the time  integrated spectrum of each GRB (as
from Tab.~1 in GGL04).  No published spectrum was found for GRB 011121
(detected by $Beppo$SAX) and this is no further considered.

$E_{\rm peak}$  versus $L_{\rm iso}$ for  the these GRBs  are shown in
Fig.~1 (blue symbols).  We omit  upper/lower limits except for the two
X--ray  Flashes with measured  $z$.  Note  that the  underluminous GRB
980425, associated  with SN 1998bw, and GRB031203,  associated with SN
2003lw,  are  major outliers  for  both  the  Yonetoku and  the  Amati
correlations.   The  statistical  results  for  the  correlations  are
reported  in  Tab.~2, together  with  the  corresponding best  fitting
power-law   parameters    (weighting   for   the    errors   on   both
coordinates). The highly significant  correlation has slope similar to
that  found by  Y04 with  12 GRBs.   The distribution  of  the scatter
measured along the correlation (i.e.  the distances of the data points
from  the  fitting  line)  of  the   22  GRBs  is  shown  in  Fig.   2
(red--hatched  histogram).  A  Gaussian fit  (red solid  line)  to the
distribution yields a scatter comparable to  that of the 23 GRB in the
$E_{\rm peak}$--$E_{\rm iso}$  plane (Ghirlanda, Ghisellini \& Firmani
(2005) - GGF05 - and black dashed line in Fig.~2).

\begin{figure} 
\begin{center} 
\vspace{8cm} \includegraphics{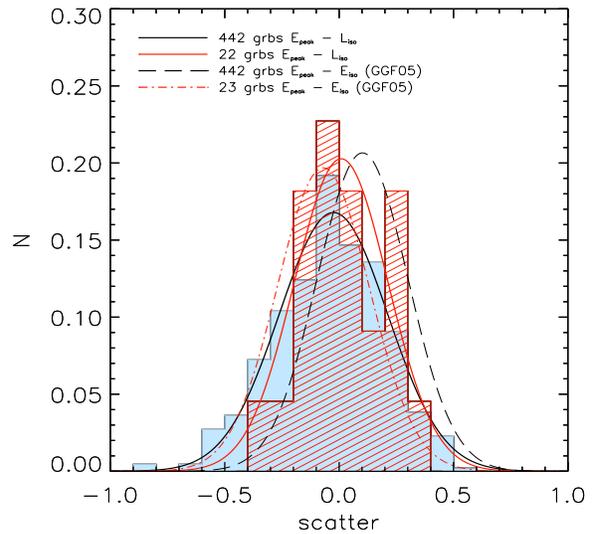}
\caption{ Distribution of the scatter of data points around their best
fit correlations.   Filled histogram: 442 GRBs with  pseudo $z$ (black
crosses in Fig.~3) around their  best powerlaw fit (solid blue line in
Fig.~3),  and Gaussian fit  to this  distribution (solid  black line).
Red hatched histogram: scatter of the 22 GRBs (blue symbols is Fig.~1)
around their best fit line (long  dashed blue line in Fig.~1), and the
best  Gaussian   fit  (red  solid   line).   Also  reported   are  the
distributions of  the scatter  of the  23 and 442  GRBs of  GGF05 with
respect  to their  correlation  in the  $E_{\rm peak}$--$E_{\rm  iso}$
plane.}
\end{center} 
\label{corr} 
\end{figure} 
For  15 out  of the  22 GRBs  listed in  Tab.~1 we  can  correct their
isotropic  luminosity   $L_{\rm  iso}$  for  the   jet  opening  angle
$\theta_{\rm   j}$  (Tab.~2   in   GGL04),  i.e.    $L_{\gamma}=L_{\rm
iso}(1-\cos\theta_{\rm j})$, with a corresponding error given by:
\begin{equation}
\left({ \sigma_{L_{\gamma}} \over L_{\gamma} }\right)^2= \left({
\sigma_{L_{\rm iso}} \over {L_{\rm iso}} }\right)^{2} +
\left({\sigma_\theta \sin\theta_{j}} \over {1-\cos\theta_{j}} \right)^2.
\end{equation}
The  red symbols  in  Fig.~1 define  the $L_{\gamma}$--$E_{\rm  peak}$
correlation.   Again all  the statistical  parameters are  reported in
Tab.~2.  The scatter  of the best fit correlation  (dashed red line in
Fig.~1) decreases with respect to  that using $L_{\rm iso}$ -- a trend
similar to  that found going  from $E_{\rm iso}$ to  $E_\gamma$ (GGL04
and  GGF05).   As  discussed  by   GGF05  in  relation  to  the  Amati
correlation, also  the scatter of the Yonetoku  correlation found here
can be interpreted  as due to the distribution  of jet opening angles.
Note, however,  that the  scatter in the  $E_{\rm peak}$--$L_{\gamma}$
correlation is  larger than that of the  Ghirlanda correlation $E_{\rm
peak}$--$E_{\gamma}$.  This fact will be discussed in Sec. 5.

%
%

%
\begin{figure} 
\begin{center} 
\vspace{8cm}   \includegraphics{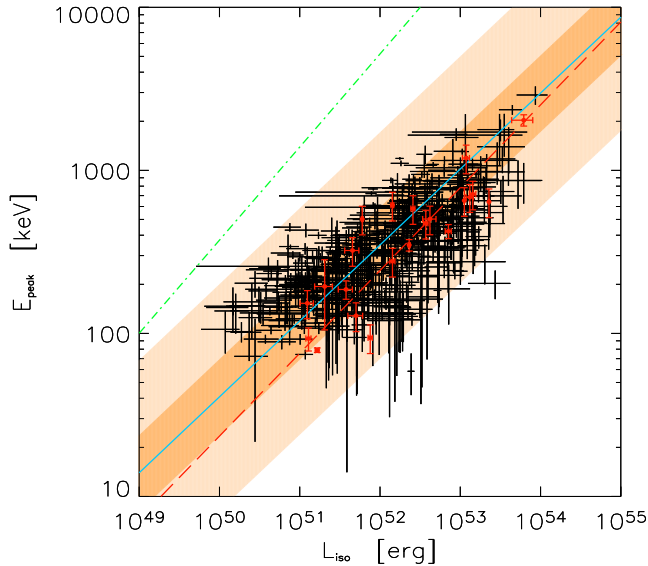}
\caption{ Rest frame peak energy $E_{\rm peak}$ versus peak luminosity
$L_{\rm  iso}$. Samples:  442 GRBs  with pseudo  $z$ defined  in GGF05
(black crosses); 22  GRBs of Tab.~1 (red symbols).  Fits: the powerlaw
fit to the 442 black crosses  (22 red symbols) is represented as solid
blue line (dashed red line). The shaded regions show the 1$\sigma$ and
$3\sigma$ width  of the dispersion  of the black crosses  around their
best fit.}
\end{center} 
\label{corr} 
\end{figure} 

\section{The pseudo redshifts sample} 

The original Amati  correlation was found with 9  $Beppo$SAX GRBs with
known $z$.  Through a redshift independent test, Nakar \& Piran (2004)
and  Band \& Preece  (2005) claimed  that the  larger BATSE  sample is
inconsistent (at 40\% and  88\% level, respectively) with the original
Amati correlation.  However, GGF05 (see also GGL04) have confirmed the
above correlation (but finding a  larger scatter) using a sample of 23
bursts with  measured $z$ as well  as using a sample  of hundreds GRBs
with  pseudo  $z$.   An   even  more  general  conclusion,  i.e.   the
consistency  of the  above  correlations with  the  entire BATSE  long
bursts sample, has been derived by Bosnjak et al. (2005).

The   same  test   of  GGF05   can  be   performed  for   the  $E_{\rm
peak}$--$L_{\rm iso}$ correlation found  in Sec. 2.  More importantly,
it is worth to investigate if  its scatter and slope change using this
much larger sample.   To this aim we consider  the same sample defined
in GGF05, which comprises 442 GRBs with pseudo $z$ [estimated by BNB04
through  the lag--luminosity relation]  and known  peak energy  of the
{\it  time integrated  spectrum}  (found by  Y04).   Since the  photon
spectral indices  are not  given in Y04,  in order to  compute $L_{\rm
iso}$ we assume typical values, i.e.  $\alpha=-0.8$, $\beta=-2.5$ (see
e.g. Preece et al. 2000).

In Fig.~3  we show these  442 GRBs (black  crosses) in the  rest frame
$E_{\rm peak}$  vs $L_{\rm  iso}$ plane,  and in Fig.   2 we  show the
distribution of the scatter of  the 442 points around this correlation
(blue histogram in Fig. 2) together with its Gaussian fit (black solid
line  in Fig.  2).  This scatter  is  only slightly  larger than  that
defined by the  22 GRBs with measured $z$  (red--hatched histogram and
red  solid line  in Fig.  2),  and the  slope of  the correlations  is
similar (see Tab. 2).

For comparison  with what  found for the  Amati correlation  by GGF05,
Fig.~2  also   reports  the  scatter  distribution   for  the  $E_{\rm
peak}$--$E_{\rm iso}$  relation with  the same samples  of 23  and 442
GRBs (dot dashed red and black dashed line, respectively).  As already
mentioned, the  scatter of the  Yonetoku correlation for the  442 GRBs
can  be interpreted  as due  to the  distribution of  the  jet opening
angles.   Assuming that  the $E_{\rm  peak}$--$L_{\gamma}$ correlation
has a smaller  scatter than the Yonetoku one, we  can estimate the jet
opening angle distribution for the  442 GRBs: we find a lognormal with
a peak  at $\theta\sim  5^\circ$, i.e. consistent  with that  found in
GGF05.

\section{The $E_{\rm peak}^{\rm p}$--$L_{\rm iso}^{\rm p}$ correlation}
 
To the aim of investigating the spectral correlations at the peak of
the prompt emission, the most correct approach would be to analyze the
spectrum of each GRB, time resolved at the burst peak.  This would
allow to derive a peak spectral energy $E_{\rm peak}^{\rm p}$ and a
luminosity (from the spectrum at the peak of the burst) 
$L_{\rm iso}^{\rm p}$ which, in general, might be different 
from the analogous integrated quantities.

Given that the GRBs listed in Tab.~1 were detected by different
satellites and that the data are public only for BATSE, we can
investigate the $E_{\rm peak}^{\rm p}$--$L_{\rm iso}^{\rm p}$
correlation only with the sample of 442 GRBs with pseudo $z$.
For these, in fact, Mallozzi et al. (1998) provide the spectral
parameters of the peak spectrum, derived by integrating the GRB signal
for $\sim$ 2 sec around the light curve peak, and BNB04 report the
peak flux corresponding to the same peak spectrum.

The sample  of 442  GRBs allows a  direct comparison with  the results
obtained  in GGF05.   However, we  here exclude  a few  bursts because
either their $E_{\rm  peak}^{\rm p}$ is below the  BATSE $\sim 30$ keV
energy  threshold   (7  cases)  or  $E_{\rm  peak}^{\rm   p}$  is  not
constrained by the spectral fit (11 cases with $\beta>-2$).
\begin{figure} 
\begin{center} 
\vspace{8cm} \includegraphics{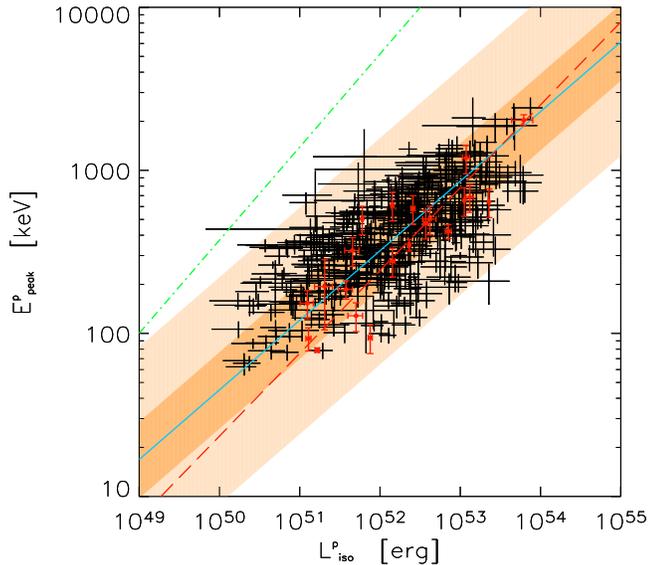}
\caption{
Rest frame peak energy $E_{\rm peak}^{\rm p}$ versus peak
luminosity $L_{\rm iso}^{\rm p}$ for the sample of 424 GRBs with
pseudo $z$.  The red symbols represent the 22 GRBs of Tab.~1 (reported
for comparison although their peak energy and luminosity are derived
assuming their time average spectrum, see Sec.~2).  The solid blue
line (dashed red line) is the powerlaw fit to the 442 black crosses
(22 red symbols).  The shaded regions represent the 1$\sigma$ and
$3\sigma$ width of the dispersion of the black crosses around their
best fit.}
\end{center} 
\label{corr} 
\end{figure} 
We report  in Fig.~4 the remaining  424 GRBs with  pseudo-$z$. Also in
this case we find a  strong correlation with a scatter consistent with
that obtained adopting $L_{\rm  iso}$ (Sec.~3), and a slightly flatter
slope. In other words, the peak  energy and luminosity at the peak are
correlated, but the correlation  is not significantly tighter than the
Yonetoku correlation (see Tab. 2).

\section{The origin of the scatter of the $E_{\rm peak}$--$L_{\gamma}$ correlation}

In Sec.~2 we have derived  the equivalent of the Ghirlanda correlation
with  the peak luminosity,  i.e.  $E_{\rm  peak}$--$L_{\gamma}$.  This
correlation (red symbols in Fig.~1)  has a scatter a factor 1.7 larger
than  that of  the Ghirlanda  correlation (for  the same  GRBs).  This
implies  that time  integrated quantities  correlate better  than time
resolved (``instantaneous") ones at the peak of the emission.

One possible reason can be  envisaged by comparing the peak luminosity
$L_{\rm iso}$ as a function  of the isotropic energy $E_{\rm iso}$ for
the  sample  of  15  GRBs  of  measured  $z$  and  jet  opening  angle
(Fig.~5). The  two quantities are correlated, but  with a considerable
scatter,  which   is  comparable  to   the  scatter  of   the  $E_{\rm
peak}$--$L_{\gamma}$ correlation: in the insert of Fig. 5 the Gaussian
fits to the scatter  distributions of the $L_{\rm iso}$--$E_{\rm iso}$
($\sigma=0.33$,  black  line)  and  the  $E_{\rm  peak}$--$L_{\gamma}$
correlation ($\sigma=0.31$, red line)  are reported. Note that in this
case  the scatter  corresponds  to the  `horizontal'  distance of  the
points from the fitting line.

The smaller scatter  of the Ghirlanda correlation with  respect to the
$E_{\rm peak}$--$L_{\gamma}$ one can be then ascribed to the fact that
$L_\gamma$   has  a   larger  spread   with  respect   to  $E_\gamma$.
Interestingly,  this  might be  related  to  $E_\gamma$  being a  time
integrated quantity,  better representing the total  kinetic energy of
the fireball, which is the  quantity involved in the derivation of the
jet opening angle (e.g. Sari, Piran \& Halpern 1999). On the contrary,
$L_\gamma$ might be more subject to local and temporal fluctuations in
the fireball, not representative of the global energetics.

\begin{figure} 
\begin{center} 
\vspace{8cm}    
\includegraphics{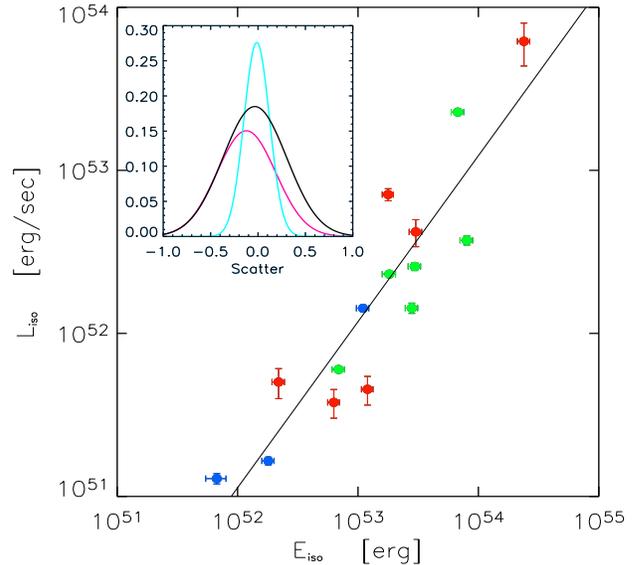}
\caption{  Peak isotropic  luminosity $L_{\rm  iso}$  versus isotropic
energy $E_{\rm iso}$ for the 15  GRBs with measured $z$ and jet angle.
Different  colors  represent  different  $z$ ranges:  $z<0.7$  (blue),
$0.7<z<1.55$ (green), $z>1.55$ (red).  The solid line (with slope 1.0)
shows the  least square fit.  The  insert reports the  Gaussian fit to
the scatter of  the data points around this  correlation (black line),
compared to  the scatter (red  line) of the  15 GRBs with  known angle
(red  points  in   Fig.~1)  around  the  $E_{\rm  peak}$--$L_{\gamma}$
correlation.  Also represented (blue line)  is the scatter of the same
15  points  in the  $E_{\rm  peak}$--$E_{\rm  iso}$  plane around  the
Ghirlanda correlation.}
\end{center} 
\label{corr} 
\end{figure} 

\begin{table*} 
\centering
\begin{tabular}{@{}lllllll@{}} 
\hline   
GRB	&		&	$\Phi_{\rm peak}$	&	$\Phi_{\rm peak}$		&	band (keV)	&       $\Phi$(1-10$^4$ keV) &      $L_{iso}$ \\
&               &       phot/cm$^2$ sec &       erg/cm$^2$ sec          &                       &       erg/cm$^2$ sec  &              erg/sec   \\  
\hline  
970228	&	S(1)	&	...		&	(3.7$\pm$0.8)E-6	&	40-700		&	7.4E-6	&	(1.6$\pm$0.3)E52\\
970828	& 	B(2) 	&	...		&	(5.93$\pm$0.34)E-6	&	30-10$^4$     	&	5.5E-6	&       (2.6$\pm$0.2)E52\\
971214	&	S(1)	&	...		&	(0.68$\pm$0.07)E-6	&	40-700		&	1.2E-6	& 	(1.3$\pm$0.1)E53\\
980425	&	B(3)	&	0.4$\pm$0.1 	&	...			&	50-300		&	1.7E-7	&	(2.7$\pm$0.35)E46\\	
980613	&	S(1)	&	...		&	(0.16$\pm$0.04)E-6	&	40-700		&	3.1E-7	&	(2.0$\pm$0.51)E51\\
980703	&	B(3)	&	2.39$\pm$0.06	&	...	        	&	50-300	     	&	1.3E-6	&       (6.0$\pm$0.2)E51\\
990123	&	S(1)	&	...		&	(17.0$\pm$5.0)E-6	&	40-700	     	&	3.7E-5	&	(6.2$\pm$1.8)E53\\
990506	&	B(3)	&	18.58$\pm$0.13  &	...			&	30-500		&	1.1E-5	&	(1.1$\pm$0.6)E53\\	
990510	&	S(1)	&	...		&	(2.47$\pm$0.21)E-6	&	40-700	     	& 	4.1E-6  &	(7.1$\pm$0.6)E52\\
990705	&	S(1)	&	...		&	(3.7$\pm$0.1)E-6	&	40-700	     	&	6.7E-6	&	(2.3$\pm$0.1)E52\\
990712	&	S(4)	&	4.1$\pm$0.3	&	...			&	40-700		&	1.9E-6	&	(1.3$\pm$0.1)E51\\
991216	&	B(3)	&	67.5$\pm$0.2	&	...			&	50-300		&	4.2E-5	&	(2.27$\pm$0.01)E53\\	
000131  &       B(3)    &       ...             &       ...                     &       ...             &       7.5E-7  &       (1.43$\pm$0.16)E53\\
000214	&	S(1)	&	...		&	(4.0$\pm$0.2)E-6	&	40-700		&	$>$9.9E-6&	$>$6.2E51\\
000911	&	I(5)	&	...		&	(2.0$\pm$0.2)E-5 	&	15-8000		&	2.0E-5	&	(1.2$\pm$0.2)E53\\   	
010222	&	S(1)	&	...		&	(8.6$\pm$0.2)E-6	&	40-700		&	$>$2.3E-5&	$>$3.2E53\\
010921	&	H(7)	&	3.19$\pm$0.3	&	...			&	50-300		&	1.7E-6	&	(1.2$\pm$0.2)E51\\
011211	&	S(6)	&	...		&	(5.0$\pm$1.0)E-8 	&	40-700		&	1.1E-7	&	(3.8$\pm$0.8)E51\\
020124	&	H(7)	&	9.38$\pm$1.77	&	...			&	2-400		&	4.6E-7	&	(4.2$\pm$0.8)E52\\
020405	&	I(5)	&	...		&	(5.0$\pm$0.2)E-6	&	15-2000		&	6.8E-6	&	(1.4$\pm$0.1)E52\\
020813	&	H(7)	&	32.31$\pm$2.07	&	...			&	2-400		&	4.1E-6	&	(3.7$\pm$0.2)E52\\
020903X	&	H(7)	&	2.78$\pm$0.67	&	...			&	2-400		&     	9.1E-9	&	$<$2.4E51\\
021211	&	H(7)	&	29.97$\pm$1.74	&	...			&	2-400	     	&	1.4E-6	&	(7.6$\pm$0.5)E51\\
030226	&	H(7)	&	...		&	(1.2$\pm$0.2)E-7	&	30-400		&	1.6E-7	&	(4.5$\pm$1.0)E51\\	
030328	&	H(7)	&	11.64$\pm$0.85	&	...			&	2-400		&	9.7E-7	&	(1.4$\pm$0.1)E52\\
030329	&	H(7)	&	450.88$\pm$24.68&	...			&	2-400		&	2.1E-5  &	(1.7$\pm$0.1)E51\\
030429	&	H(7)	&	3.79$\pm$0.79	&	...			&	2-400		&	8.7E-8	&	(5.0$\pm$1.0)E51\\	
030723X	&	H(7)    &	2.10$\pm$0.41	&	...			&	2-400		&	3.7E-8	&	$<$1.2E51\\
031203  &       I(8)    &       ...             &      2.4E-7                   &       20-200          &       $>$9.6E-9 &     $>$2.8E47\\
\hline 
\end{tabular} 
\caption{Peak fluxes and bolometric luminosities for GRBs with
measured $z$ (in GGL04).  S={\it Beppo}SAX, B=Batse,
I=Integral, H=Hete--II.  
Photon peak fluxes or energy peak fluxes with references (col.~2) 
and corresponding observed energy band (col.~5).  
References: 
(1) Amati et al. 2002; 
(2) Yonetoku et al. 2004; 
(3) 4th Batse catalog; 
(4) Frontera et al. 2001; 
(5) Price et al.  2002; 
(6) Piro et al. 2004; 
(7) Sakamoto et al. 2004. 
(8) Sazonov et al. 2004.}
\end{table*}

\begin{table*} 
\centering
\begin{tabular}{@{}lllllllllll@{}} 
\hline   
Correlation &$N$ &$r_{s}$ &$P$& $r_{z}$ &$A$  &$S_o$ &$\delta$      &$\chi^{2}$/d.o.f. &$\mu$ &$\sigma$ \\
\hline 
$E_{\rm peak}$--$L_{\rm iso}$ & 22 
  &0.87 &$8.5\times 10^{-8}$ & 0.83  &$2.00\pm0.05$ &$7.4\times 10^{51}$ &$0.50\pm 0.02$ & 127/20  &~~0.01   &0.20 \\
$E_{\rm peak}$--$L_{\gamma}$   &  15 
  &0.88 &$1.6\times10^{-5}$ & 0.85   &$2.56\pm0.12$ &$4.3\times 10^{49}$ &$0.57\pm 0.03$ &49.6/13  &~~0.04    &0.17\\
$E_{\rm peak}$--$E_{\gamma}$  &  15   
  &0.95 & $2.8\times10^{-8}$& 0.94   &$2.62\pm0.15$ &$4.2\times10^{50}$  &$0.70\pm0.05$  &16.5/13  &--0.07    &0.10\\
$E_{\rm peak}$--$L_{\rm iso}$ &  442 
  &0.71  &$1.6\times 10^{-69}$& 0.7  &$4.36\pm 0.04$&$1.6\times 10^{52}$&$0.47\pm 0.01$ &4171/440 &--0.02 &0.23\\
$E_{\rm peak}^{\rm p}$--$L_{\rm iso}^{\rm p}$ &  424 
  &0.66 &$5.1\times 10^{-65}$& 0.65  &$3.80\pm 0.06$&$1.6\times 10^{52}$&$0.43\pm 0.01$ &4154/422 &~~0.05   &0.23\\
\hline 
\end{tabular} 
\caption{Statistical results.
$N$: Number of objects;
$r_{s}$: Spearman correlation coefficient;
$P$: Chance Probability; 
$r_{z}$: Partial correlation coefficient subtracting the effect of  $z$ (e.g. Wall \& Jenkins 2003);
$A$, $S_o$ and $\Omega$: Linear Fit Normalization, Scaling and Slope, 
    i.e.: $(E_{\rm peak}/100$ keV) = $A (S/S_o)^\delta$; 
$\mu$ and $\sigma$: Gaussian fit parameters.
}
\end{table*} 
 
\section{Conclusions} 

We  derived the  $E_{\rm  peak}$--$L_{\rm iso}$  correlation with  the
current largest  available sample of 22  GRBs with known  $z$ and well
determined spectral properties (GGL04).   This correlation has a slope
0.51, i.e.  similar  to that proposed by Y04,  although its scatter is
much larger than  they originally found with 12  GRBs.  The scatter is
instead  comparable with what  GGF05 found  for the  Amati correlation
using the same sample of GRBs.  Using the 442 GRBs with pseudo--$z$ we
still  find a strong  correlation, with  similar scatter  and slightly
flatter slope than those found with the 22 GRB of measured $z$.

We  then  considered the  robustness  of  correlations for  quantities
calculated  at  the  peak  of  the  emission,  with  respect  to  time
integrated  properties. The former,  in particular  $E_{\rm peak}^{\rm
p}$ vs $L_{\rm  iso}^{\rm p}$, results in a  correlation equally tight
to that involving integrated quantities.

Correcting  $L_{\rm   iso}$  for   collimation,  we  found   that  the
corresponding  $E_{\rm peak}$--$L_{\gamma}$  correlation  has a  slope
flatter  than the  Ghirlanda correlation  (0.57 vs  0.7) and  a larger
scatter (0.17  vs 0.1).  The larger  scatter might be  ascribed to the
fact that  the peak  luminosity is less  representative of  the global
energetics of  the burst,  which in turn  is adopted to  represent the
total kinetic energy of the fireball and thus estimate the jet opening
angle (e.g.  Sari et al.  1999).

\vspace{-0.5cm}
\section*{Acknowledgments} 
The   Italian  MIUR  and   INAF  are   acknowledged  for   funding  by
G. Ghirlanda, G. Ghisellini (Cofin grant 2003020775\_002), AC and ZB.

\end{document}